\documentclass[twocolumn,
aps,nofootinbib,showpacs,showkeys, 
tightenlines,preprintnumbers,]{revtex4}

\usepackage{epsf,epsfig,subfigure,graphicx,amsmath,amssymb}
\usepackage{color}
\newcommand{\dis}[1]{\begin{equation}\begin{split}#1\end{split}\end{equation}}

\newcommand{\gev}{\,\textrm{GeV}}

\newcommand{\eV}{\,\mathrm{eV}}
\newcommand{\Mp}{M_{\rm P}}
\newcommand{\Mpt}{$M_{\rm P}$}
\newcommand{\Mg}{$M_{\rm GUT}$}
\newcommand{\Uone}{U(1)$_{\rm gl}$}

\newcommand{\UPQ}{U(1)$_{\rm PQ}$}
\newcommand{\tut}{$t_{\rm U}$}
\newcommand{\tu}{t_{\rm U}}

\newcommand{\ie}{{\it i.e.~}}
\newcommand{\etal}{{\it et al.}\,}

\begin{document}

\title{ \Large\bf Bosonic Coherent Motions in the Universe}

\author{Jihn E.  Kim}
\address{
 Department of Physics, Seoul National University, Seoul 151-747,  Korea, and\\
  Department of Physics, Kyung Hee University, Seoul 130-701,  Korea
}

\author{Yannis Semertzidis}
\address{Center for Axion and Precision Physics Research (IBS),
  Daejeon 305-701,  Korea, and\\
 Department of Physics, KAIST, Daejeon 305-701,  Korea}

\author{Shinji Tsujikawa}
\address{Department of Physics, Faculty of Science, Tokyo University of Science,
1-3 Kagurazaka, Shinjuku, Tokyo 162-8601, Japan}

\begin{abstract}
We review the role of fundamental spin-0 bosons as bosonic coherent
motion (BCM) in the Universe. The fundamental spin-0 bosons have the potential
to account for the baryon number generation, cold dark matter (CDM) via BCM,
inflation, and dark energy. Among these, we pay particular attention to
the CDM possibility because it can be experimentally tested with
the current experimental techniques.
We also comment on the panoply of the other roles of spin-0
bosons--such as those for cosmic accelerations at early and late times.

\end{abstract}

\maketitle

\section{Introduction}

Recent cosmological observations \cite{Planck13,WMAP9} confirm the eight decade
old  Zwicky's proposal  \cite{ZwickyF33} that the Universe contains a large amount
of dark matter (DM). The DM profile has been measured accurately enough to
pinpoint DM to ``it is cold dark matter (CDM) \cite{Planck13}."
The bosonic coherent motion (BCM) can be CDM \cite{Preskill83,Abbott83,Dine83} if
the coherent-boson lifetime is long enough to have survived until
now \cite{KSVZ,Shifman80,DFSZ,Dine81}.
The axion proposed to solve the strong
CP problem \cite{KimPRP,Cheng88,Pecceirev,KimRMP}
is fitting to this BCM scenario \cite{RevAxionsCDM}.
The BCM is one of many possibilities of CDM scenarios \cite{RevAxionsCDM}
which accounts for only 27\,\% in the energy pie of the Universe.
The dominant portion, 68\,\%, in the energy pie is the homogeneous energy density,
at least up to the $10^3$ Mpc scale, which is usually referred to dark energy.
Dark energy (DE) being homogeneous cannot be accounted for by corpuscular
particles but may be accountable by the cosmological constant or
by some vacuum expectation value (VEV) of spin-0
boson(s) \cite{Sahni,Pratra,Paddy,DErev,Silvestri09,Caldwell09,Tsuji10}.
The visible particles (mostly atoms in the energy count) constitute
only 5\,\% in the energy pie.

If we accept the Big Bang cosmology from the earliest possible time, $10^{-43}$ s,
the success of the Standard Model of particle physics is based on the
assumption of very tiny DE of order less than $10^{-46}\,\gev^4$
because the age of the Universe is very long $\simeq 13.8\,{\rm Gy}$
\cite{Planck13,WMAP9}.
So, the DE problem or the theoretical cosmological constant  problem
\cite{WeinbergCC} is not
only the problem in cosmology but also a problem in particle physics.
Out of despair, many adopt the anthropic scenario for the cosmological constant problem
\cite{WeinbergAnth,Tegmark06,ZeeEgr13}. For the anthropic solution to work, the cosmological constant must be a free undetermined parameter in particle physics, as integration constants of Refs. \cite{Hawking84,KKL00,KimCC}. In a deterministic theory such as
in string theory, possible cosmological constants must be allowed near 0
for our Universe to have adopted one of these, which is the reason
trying to have as many as $10^{120}$ models,
to pack the vacua with separation between them by
$(10^{-3\,}\eV)^4$, from string theory \cite{ZeeEgr13}.
But, all those  $10^{120}$ vacua must allow three family SMs,
and satisfy the known SM phenomena such as the GUT scale weak
mixing angle $\sin^2\theta_W=3/8$ \cite{Kim81,Luo91,Kim03}, etc.
But, we have only a handful of minimal supersymmetric Standard Models
from string theory satisfying the requirements \cite{stringSM,KimKyae07,HuhKK09,Lebedev07,Nilles06}.
Or, a Standard Model solution with DE$\simeq 10^{-47\,}{\rm GeV}^4$ has to be found so that the anthropic argument chooses it. This search seems more difficult than finding a vanishing cosmological constant solution theoretically.
At present, we can say that the anthopic solution in string theory has not worked out yet. Therefore, in the  Standard Model and in its supergravity extension, it is fair to say that the cosmological constant is assumed to be zero.

By observing the luminosities of Type-I supernovae \cite{Perlmutter99,Riess98},
the recent acceleration of the Universe has been established.
So, explaining the DE scale of $10^{-47}\,\gev^4- 10^{-46}\,\gev^4$
became an important topic \cite{DErev,Silvestri09,Caldwell09}.
In   \textbf{Table\,\ref{Tab:01}}, we list several ideas proposed to account for
this recent acceleration of the Universe.
Both the high scale inflation \cite{inflationold,Infnew1,InfNew2}
and the recent acceleration  \cite{Perlmutter99,Riess98} in the Universe
are based on the assumption  of vanishing cosmological constant.

\begin{figure}[b!]
\begin{center}
\begin{tabular}{c}
\includegraphics[width=8.5cm]{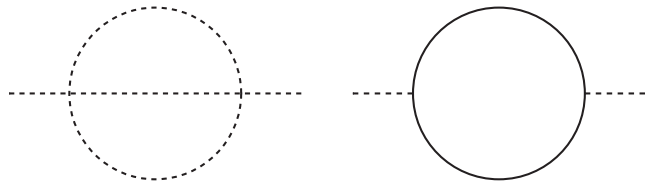}
\end{tabular}
\end{center}
\caption{Diagrams leading to dimension 2 interactions
with dimension 4 coupling at each vertex.
}\label{DilBreaking}
\end{figure}

To determine the VEV of a scalar field, say $\phi$, one must consider
all the allowed effective terms at low energy. At each interaction point,
suitable symmetry requirements must be satisfied.
A typical mass scale of $\phi$ is given by the effective mass term
$m^2|\phi|^2$. In Fig. ~\ref{DilBreaking}, we consider only two diagrams
with the dimension 4 ($d=4$) couplings. If each $d=4$ vertex of
Fig.~\ref{DilBreaking} satisfies the global phase symmetry,
the two-loop and one-loop mass terms do not break the global symmetry.
On the other hand, each $d=4$ vertex satisfies the dilaton
symmetry (requiring just $d=4$ couplings) but the diagrams of Fig. \ref{DilBreaking}
are $d=2$ terms which of course break the dilaton symmetry.
One well-known model breaking the dilaton symmetry at the one-loop
quantum level, including the $d\ge 6$ terms, is the Coleman-Weinberg
model \cite{ColemanW73}. Therefore, it is not likely that a consistent
calculation of a small DE scale can be performed by introducing the
dilaton symmetry. However, some global phase symmetry may be suitable for this.

In Sec. \ref{sec:spin0}, we present the focus points of this
review: the BCM scenarios and the axion detection experiments.
In Sec. \ref{sec:CC}, we point out the difficulty of obtaining zero cosmological constant theoretically.
In Sec. \ref{sec:inflation}, we mini-review the inflationary cosmology, in particular in view of the recent BICEP2 data.
In Sec. \ref{sec:discussion}, we discuss the subject of this review: why the role of {\it fundamental} spin-0 particles are important in cosmology.

\section{Spin-0 boson filling the Universe}\label{sec:spin0}

After the discovery of a fundamental spin-0 scalar particle (the Brout-Englert-Higgs   boson)
at the LHC, it  is timely to study the roles of fundamental spin-0 bosons in the Universe.
It is very interesting to note that fundamental spin-0 bosons have been employed to account
for the mothers of atoms ({\it i.e.} baryon number generation via the Affleck-Dine mechanism \cite{AffDine}), CDM via BCM \cite{Preskill83}, DE via a transient cosmological constant  \cite{QuintAx03,KimQ99,Kimq00,KimNill09,Choi00},
and even the vacuum energy needed for the high scale inflation \cite{inflationold,Infnew1,InfNew2}.
Among these, we focus on CDM via BCM in this review because similar ideas can be applicable to DE and inflation models. Another attractive point discussing CDM via BCM is that it can be experimentally proved in the near future \cite{Yannis14}.

\begin{widetext}

\begin{table}[t!]
\begin{center}
\begin{tabular}{l|llll}\hline
Ideas & Description [scalar\,S or pseudoscalar\,P] & ~~Disc. sym. & Fine-tune & From string\\
\hline
MOND$^{\,\rm a}$ &Change Newtonian gravity. [No boson]  & ~~Irrelevant & Yes& Not yet \\[0.2em]
Anthropic principle & Out of many possible vacua, only those    & ~~Irrelevant & Irrelevant & Not yet$^{b}$\\
           & suitable for age\,$>t{\rm _U}$ survived. [S or P] & & &\\[0.2em]
Quintessence & With a runaway $V\propto 1/\phi^n$ ($n>0$). [S] & ~~No & Yes$^{\rm c}$ & Not yet\\[0.2em]
Dilaton & P-Gold. boson from dilaton sym. [S] & ~~No& Yes$^{\rm d}$ & Not yet\\[0.2em]
U(1)$_{\rm DE}$ Goldstone& P-Gold. boson from U(1)$_{\rm DE}$ sym. [P] & ~~Yes & No &Yes$^{\rm e}$ \\ \hline
\end{tabular}
\end{center}
\caption{Typical DE models with a few  pseudo-Goldstone bosons originating from global symmetries.
$^{\rm a}$ Refs.~\cite{MOND,Becken}, $^{\rm b}$ Ref.~\cite{ZeeEgr13}, $^{\rm c}$
Ref.~\cite{Stenhardt98}, $^{\rm d}$ Ref. \cite{Wetterich}, $^{\rm e}$ Ref.~\cite{KimNilles14}.
}\label{Tab:01}
\end{table}

\end{widetext}

We are familiar with the ether idea of the late 19th Century, filling out the Universe.
The VEV idea of spin-0 particles used for breaking global symmetries \cite{Goldstone61}
and gauge symmetries \cite{Higgs64} is a kind of ether.
If a scalar field $\phi$ has a universal value over the entire Universe,
any operation of the type `Poincare transformation' does not notice a change.
Thus, the VEV of a scalar field,  $\langle\phi\rangle$, respects the Poincare symmetry.
But, if $\phi$ is a complex field, then the VEV breaks the phase transformation
symmetry, {\it i.e.} breaks a global phase symmetry \cite{Goldstone61}.
Even though the Brout-Englert-Higgs mechanism \cite{BE64,Higgs64,GHK64}
for breaking gauge symmetries is not a monopoly of spin-0 particles \cite{Techni,WeinTechni},
now the role of spin-0 particles becomes more important, especially after
a hint of large tensor-to-scalar ratio $r$, based on  the BICEP2 observation \cite{Larger}.

Let us denote  scalar and pseudoscalar particles as $s$ and $a$, respectively.
Scalar particles transform under the parity operation as
$P: s({\bf x})  \to +s(-{\bf x})$, and pseudoscalar particles transform
as $P: a({\bf x}) \to -a(-{\bf x}) $.
If they are components of a complex field, it is usually represented as the
radial and phase fields, respectively, $\phi=s e^{ia/f}$ where $f$ is a mass parameter.
Thus, the complex field transforms under parity as
$P: \phi({\bf x}) \to\phi^*(-{\bf x}) $.
Any pseudoscalar field represented as a phase can be represented by
an angle field with the angle defined in the range $[0,2N\pi)$, where $N$
is the domain-wall number. A Goldstone boson arising from breaking a
global phase symmetry by the VEV $v$ is a pseudoscalar
field $a$ defined as
\dis{
\langle \phi\rangle=\frac{v+s}{\sqrt2}\, e^{ia/f},~~~
\langle s\rangle=0,~~~\langle a\rangle=[0, 2N\pi f).
}

\subsection{Cosmology with BCM}\label{subsec:spin0cosm}

\begin{figure}
\begin{center}
\begin{tabular}{c}
\includegraphics[width=7cm]{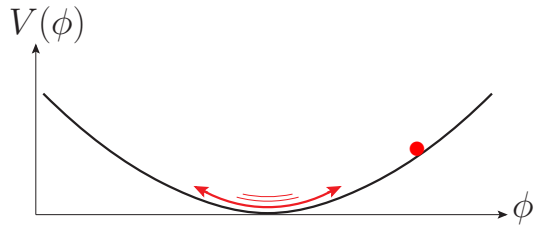}
\end{tabular}
\end{center}
\caption{ After $t_1$, the BCM potential of $\langle\phi\rangle$
at the red bullet oscillates around the minimum.}\label{BCMcosmology}
\end{figure}

On the flat Friedmann-Lemaitre-Robertson-Walker
cosmological background space described by the line element
$ds^2=-dt^2+a^2(t)\delta_{ij}dx^i dx^j$,
the evolution of the classical scalar field $\phi$,
(\ie the evolution of the VEV of $\phi$), is given by
\dis{
\frac{d^2}{dt^2}\langle \phi\rangle+
3 H\frac{d}{dt}\langle \phi\rangle+V'(\langle \phi\rangle)=0\,,\label{eq:vev}
}
where $H=\dot{a}/a$ is the Hubble parameter and
$V'=(d/d\langle \phi\rangle)V$ is a derivative of the
potential $V$ (a dot represents a derivative with
respect to the cosmic time $t$).
With a discrete symmetry $\phi \to -\phi$, the leading term of $V'$ is
the mass term $m^2 \langle \phi \rangle$.
When $\langle \phi \rangle$ moves very slowly, we can neglect the
second derivative term, and the evolution equation gives
$3H \dot\phi \simeq -m^2 \phi$. $\langle \phi \rangle$ starts to
change rapidly when $H$ becomes small enough to satisfy
$3H\simeq m$. After this condition is met, $\langle \phi \rangle$
oscillates rapidly, as shown in Fig.~\ref{BCMcosmology},
which is interpreted as the BCM of $\phi$.

As mentioned above, the VEV $ \langle \phi \rangle$ is assumed to be the same
over the whole Universe for the Poincare invariance,
otherwise the invariance is broken. In the Universe, this homogeneity is subtly broken.
The inflation manages different scales of quantum fluctuations enter the horizon
at different scales, basically breaking the homogeneity.
A given scale condenses gravitationally.
The VEV in that scale evolves according to
Eq.~(\ref{eq:vev}), and describes the BCM of $\langle\phi
\rangle$.

From the Friedmann equation we have $3H^2\Mp^2=\rho$,
where $\rho$ is the energy density of the Universe and
$\Mp$ is the reduced Planck mass ($\Mp=2.4 \times 10^{18}$ GeV).
Denoting the time at the onset of oscillations of $\langle\phi\rangle$
as $t_1$, the condition for determining $t_1$ is
\dis{
\sqrt{\frac{3\rho(t_1)}{M_{\rm P}^2}}=m(t_1).
}
These oscillations are equivalent to a gas of $\phi$ particles of low-momentum.
This kind of spin-0 particle coherent motion was first discussed in
Ref.~\cite{Preskill83,Abbott83,Dine83}  for the case $\phi={\rm axion}$.
It is known that the BCM behaves like CDM because  of the low-momentum.
Thus, the number and energy densities are given by
\dis{
n=m\langle \phi\rangle^2\,,\quad
\rho= m^2\langle \phi\rangle^2\,.
}

We denote the current age of the Universe as \tut.
Depending on $t_1$ and \tut, we can classify BCMs as
\begin{itemize}

\item \indent  {\bf BCM}:
If $t_1<\tu$, the currently oscillating vacuum $\langle\phi\rangle$ is BCM.
The BCM can be classified into the following two sub-categories.
\begin{itemize}
\item[] $\star$ {\bf BCM1}:
The lifetime of $\phi$ is long enough, $\tau_{\phi}>\tu$.
Then, the oscillating BCM contributes to the CDM amount.
The QCD axion belongs here.
\item[] $\star$  {\bf BCM2}: The lifetime of $\phi$ is short, $\tau_{\phi}<\tu$.
Then, all $\phi$ quanta decayed already, producing SM particles.
The inflaton with $\tau_\phi\sim 10^{-36}$\,s belongs here and
reheating after inflation gives the beginning of the
radiation-dominated Universe.
\end{itemize}
\item\indent {\bf CCtmp}: Temporary cosmological constant.
On the other hand, if $\langle\phi\rangle$ has not oscillated yet,
then $t_1>\tu$ and $\langle\phi\rangle$ stays there now, behaves
like a cosmological constant, but it is a temporary phenomenon and will eventually become
{\bf BCM1} after $t_1$.
For this to be satisfied, the mass is around $10^{-33\,}$eV with
a trans-Planckian decay constant \cite{Carrol98}.
If $V(\langle\phi\rangle)$ describes {\bf CCtmp}, the equation of
state $w_\phi$, characterized by the field energy density
$\rho_{\phi}=\frac12 \dot{\phi}^2 +V(\phi)$ and the pressure
$P_{\phi}=\frac12 \dot{\phi}^2 -V(\phi)$, is a useful parameter,
\dis{
w_{\phi} \equiv \frac{P_{\phi}}{\rho_{\phi}}
=\frac{\frac12 \dot{\phi}^2 -V(\phi)}{\frac12 \dot{\phi}^2 +V(\phi)}\,.\label{eq:ofstate}
}
Provided that $\frac12 \dot{\phi}^2  \ll V(\phi)$, $w_{\phi}$ is close to $-1$,
behaving like the cosmological constant. In order to realize the recent acceleration,
we require the condition $w_{\phi}<-\frac13$.

\end{itemize}

\subsection{Scalar particles}

The Brout-Englert-Higgs boson is the only known fundamental scalar field.
The other scalar most widely used in particle theory is dilaton,
the scalar Goldstone boson arising from breaking the dilatonic symmetry.
The effect of dilatonic symmetry on the cosmological constant problem has been discussed
extensively in Ref.~\cite{Wetterich}.
For the solution, however, a fine-tuning is needed.
The obvious effect of a VEV of a scalar field $s$ in cosmology is the
universal constant on the right-hand-side  of the Einstein equation.

The scalar-field cosmology in the presence of a barotropic perfect fluid
was studied in 1980--90s \cite{Fujii82,Fujii90,Ford87,Wetterich,Peebles88,Ratra88,Chiba97,
Ferreira97,Paul98,CLW98}, even before the discovery of
the recent cosmic acceleration.
This was chiefly motivated by the  ``missing matter problem'' in 1980s.
In 1990, Fukugita \etal \cite{Fukugita90} tested cosmological
models against observations of the number count of faint
galaxies and showed that these data favor the Universe
with low matter density ({\it i.e.}, matter is missing).
In the abstract of their paper they stated that
``Furthermore, it is shown that the best agreement with
the data is obtained with a sizable cosmological constant,
including the case of zero curvature model as predicted
by inflation.'' In addition, it was already known in the early 1990s that
the presence of a cosmological constant can make the age of the Universe
longer such that it is consistent with the age of oldest globular clusters.

If the cosmological constant originates from a vacuum energy
appearing in particle physics, it is vastly larger than the
today's average cosmological density \cite{WeinbergCC}.
Because of this problem, people tried to construct
dynamical cosmological constant models in which the energy density of
cosmological constant varies in time,
basically belonging to a kind of {\bf CCtmp}.
For example, if we consider a dilaton field $\phi$, the cosmological constant
depends on $\phi$ by transforming the dilatonic action
to the so-called Einstein-frame action (in which the dilaton does not
have a direct coupling with the Ricci scalar) \cite{Fujii90,Wetterich}.

Exponential potentials often arise from the curvature of internal spaces
associated with the geometry of extra dimensions
(so called ``modulus'' fields) \cite{GSW,Olive90}.
Inspired by this, the exponential potential $V(s)=V_0 e^{-\lambda\, s\Mp}$
has been used, with constant parameters $V_0$ and $\lambda$.
There are two distinct fixed points on the flat  Friedmann-Lemaitre-Robertson-Walker
cosmological background space  \cite{CLW98,DErev}: (a) the scaling solution,
and (b) the scalar-field dominated solution.

For $\lambda^2>3(1+w_m)$, where $w_m$ is the equation of state
for a background fluid,
the solutions approach the scaling fixed point (a),
characterized by the field density parameter $\Omega_{s}=3(1+w_m)/\lambda^2$
and the field equation of state $w_{s}=w_m$ \cite{Peebles88,Ferreira97,CLW98}.
Even for the initial conditions where $\rho_{s}$ is larger than $\rho_m$
in the early radiation-dominated era, the field eventually enters the scaling regime in which
$\rho_{s}$ is proportional to $\rho_m$ with $\rho_{s}/\rho_m={\rm constant}<1$.
The field energy density of the scaling solution contributes to the total energy
density of the Universe, but it does not lead to the cosmic acceleration.
For $\lambda^2<3(1+w_m)$, there exists the scalar-field dominated
fixed point (b), characterized by $\Omega_{s}=1$ and $w_{s}=-1+\lambda^2/3$.
The late-time cosmic acceleration can be realized for $\lambda^2<2$.
Since in this case the point (b) is also stable, the scalar field can be the source of DE.
For $\lambda^2<2$, the scalar potential is quite shallow, so the field density in the
early Universe needs to be much smaller than the background energy density
(unlike the scaling solution discussed above).

After the discovery of the recent cosmic acceleration in 1998, the cosmological dynamics
of ``quintessence'' (a canonical scalar field responsible for DE) were studied in detail
for several different potentials \cite{Paul2,Paul3}.
One example is the inverse power-law potential
$V(s)=M^{4+n}s^{-n}$, where $M$ and $n$ are positive constants.
This potential can arise in globally supersymmetric QCD
theories \cite{Bine}\footnote{However, the scalar in this case is composite.}.
The Universe enters the stage of cosmic acceleration for the field value larger than
$s_0 \approx \Mp$. Since $V(s_0)$ is of the order of $H_0^2\Mp^2$,
one can estimate the mass scale $M$ as $M \approx 10^{-(46-19n)/(4+n)}$ GeV.
For $n=O(1)$,
this energy scale can be compatible with that appearing in particle physics.

In the presence of a perfect fluid with
the equation of state $w_m$, there exists a so-called tracker
solution for the potential $V(s)=M^{4+n}s^{-n}$.
The tracker is characterized by a common, cosmic evolutionary trajectory
that attracts solutions with a wide range of initial conditions \cite{Paul3}.
The field equation of state along the tracker is given by $w_{s}=(w_mn-2)/(n+2)$,
which corresponds to $w_{s}=-2/(n+2)>-1$ during the matter era.
The slope of the potential $\lambda=-\Mp V_{,s}/V=n\Mp/s$
gets smaller with the growth of $s$, so $w_{s}$ approaches $-1$ in the future.
The inverse power-law potential belongs to a class of freezing quintessence
models \cite{CaLi} in which the evolution of the field gradually slows down.

There is another class of quintessence models, dubbed thawing
models \cite{CaLi}, in which the field has been frozen by Hubble friction
and then it starts to evolve after the Hubble parameter drops below the field mass $m$.
In this case the field equation of state $w_{s}$ is close to $-1$ at the initial stage,
but it starts to grow at the late cosmological epoch.
The field mass $m_s$ responsible for dark energy corresponds to
$m_s \simeq 10^{-33}$ eV \cite{Carrol98}.
The representative potential of thawing models is that of
a pseudo-scalar field arising from breaking the global U(1)
symmetry (which we will explain more details in Sec.~\ref{pseudosec}).

If we consider a scalar field $\phi$ non-minimally coupled to the Ricci scalar $R$ (like dilaton),
this gives rise to a coupling with non-relativistic matter in the Einstein frame \cite{Amen}.
The fifth force induced by such a matter coupling needs to be suppressed in the solar system.
There are several ways to suppress the propagation of the fifth force in local regions of the Universe.

One is the so-called chameleon mechanism \cite{Khoury}, under which the mass of
a scalar degree of freedom is different depending on the matter densities in the
surrounding environment. If the effective mass is sufficiently large in the regions of high density,
the coupling between the field and non-relativistic matter can be suppressed by having a thin shell
inside a spherically symmetric body. In Brans-Dicke theory (including $f(R)$ gravity) \cite{Brans}
it is possible to suppress the propagation of the fifth force by designing
the field potential $V(\phi)$ appropriately \cite{Khoury2,Sawi,Capo,Mota,fRgravity1,fRgravity2}.

Another is the so-called Vainshtein mechanism \cite{Vain}, under which
nonlinear scalar-field self interactions can suppress the fifth force at short
distances even in the absence of the field potential.
The self interactions of the form $(\partial \phi)^2 \square \phi$,
which correspond to the Lagrangian of covariant Galileons \cite{Galileons},
can lead to the decoupling of the field $\phi$ from matter
within a radius much larger than the solar-system scale \cite{Nicolis,Kimura,Kase,Kase2}.

\subsection{Pseudoscalar particles}\label{pseudosec}

Most pseudoscalar particles observed so far are pseudo-Goldstone bosons.
Let $a, \Lambda$ and $f$, respectively, be a Goldstone boson from a
spontaneously-broken global U(1) symmetry, the dominant explicit
symmetry breaking mass parameter, and the decay constant. Then, the mass of $a$ is
\dis{
m_a=c_a \frac{\Lambda^2}{f}\,,
}
where $c_a$ is the number given by the explicit symmetry breaking terms.
For the QCD axion, the breaking of the U(1) symmetry
is given by the QCD anomaly and we have $c_a \Lambda^2=[Z^{1/2}/(1+Z)] f_\pi m_\pi$
with $Z=m_u/m_d$ where $f_\pi, m_\pi, m_u,m_d$ are neutral-pion decay constant,
its mass, and $u$ and $d$ quark masses \cite{KimRMP}.
If the explicit breaking term is given
by $V_{\rm br}=-(\Lambda^{4-n}\phi^n+{\rm h.c.})/2$,
then we have $m_a= (f/\Lambda)^{n/2}  (n\Lambda^2/f) $.
As shown in Fig.~\ref{DilBreaking}, the pseudo-Goldstone boson arising from
a global symmetry \Uone~ does not appear in the loops if each vertex
satisfies \Uone. But, it is known that all global symmetries are
approximate \cite{KimNilles14,Kim13worm,Diaz14,KraussW89,KimJKPS14}. Most strong explicit breaking may be from the anomaly of the type \Uone-$G$-$G$,
where $G$ is a non-Abelian gauge group.

The most waited-for pseudoscalar particle is the very light axion in the
axion window because its  discovery will confirm at least three: (1) a physical confirmation of instanton solutions of non-Abelian gauge theories \cite{Belavin75},
(2) \'{}t Hooft solution \cite{HooftSol} of the U(1) problem of QCD \cite{Weinberg75}, and (3) at least some portion of CDM in the Universe. The particle {\it axion} was first appreciated by Weinberg and Wilczek in the Ben Lee Memorial Conference in October, 1977 \cite{Cline77}, using the  Peccei-Quinn (PQ) symmetry  \cite{PQ77}.
If $G$ is QCD, the symmetry \Uone~is called the PQ
symmetry \UPQ~and the pseudo-Goldstone boson $a$ related to \UPQ~
 is called the {\it QCD axion}.
The axion is needed to understand the strong CP problem of ``Why is the neutron electric diplole moment so small even though the gluon interactions (in the presence of instanton solutions of QCD) allow a neutron-size diplole moment?'' In early days, three kinds of solutions to the strong CP problem were admitted \cite{KimPRP}: the calculable solution, the massless up quark case, and the axion solution. The calculable solutions have not provided yet an acceptable model with sufficiently small neutron electric diplole moment. The massless up quark case is not favored in the global fit \cite{KimRMP}. The remaining axion solution is checked in various cases as discussed in the next Subsect.

Field theory examples on axions with renormalizable couplings corresponding to {\bf BCM1}
are usually classified to the KSVZ and DFSZ models \cite{KSVZ,Shifman80,DFSZ,Dine81}.
But, this classification is too simple. There can be many KSVZ and DFSZ type models
with one type of quark representations \cite{Kim98}.
One may introduce many different types of quarks also for axion phenomenology.

Therefore, it is better to have a theory predicting definite PQ charges of the quarks in a full theory. The most attractive proposal along this line is the string compactification. Here,
the PQ global symmetry is determined once the compactification scheme is presented. Standard models obtained from string compactification include many quarks beyond the Standard Model spectrum, in particular numerous singlet fields. Along this line, several years ago a QCD axion including non-renormalizable terms was studied and the axion-photon-photon coupling has been calculated with an approximate \UPQ~symmetry \cite{ChoiKimKim}. Recently, an exact \UPQ~symmetry has been studied in a string compactification where the axion-photon-photon coupling has been calculated below the PQ symmetry breaking scale \cite{Kimagg14},
\dis{
c_{a\gamma\gamma} =\frac{1123}{388}-1.98\simeq 0.91\,.
}
We expect that more calculations of $c_{a\gamma\gamma} $ will be performed
in string models with the property of successful SM phenomenologies,
which will guide us where to look for the QCD axion \cite{Yannis14}.

Dark energy can be the case of {\bf CCtmp} in the above classification.
Pseudoscalar {\bf CCtmp} have been discussed already more than a decade ago
in Refs.~\cite{QuintAx03,Choi00,Nomura}.
But, a more plausible analysis, looking into the
detail of string compactification, has been presented
recently \cite{KimNilles14,KimJKPS14,KamionAxi14}.

The field mass $m_a$ responsible for dark energy corresponds to
$m_a \simeq 10^{-33}$ eV \cite{Carrol98}. Meanwhile, if the axion field is responsible
for CDM, the typical mass scale is between $10^{-5}$ eV and $10^{-2}$ eV \cite{KimRMP}.
In string theory there are many ultralight axions possibly down
to the Hubble scale $H_0=10^{-33}$ eV \cite{axiverse}.
Axions in the mass range between $10^{-28}$ eV and $10^{-18}$ eV become
non-relativistic at a later cosmological epoch relative to the standard CDM.
Such a light scalar field leads to the suppression of the CDM power spectrum on
small scales \cite{axiverse,Barbi,Wayne} (like light massive neutrinos),
so there is an observational signature for ultralight axions
if the axion potential is of the form $[1-\cos(a/f_a)]^3$ \cite{KamionAxi14}.\footnote{For this specific form, one needs fine-tunings between domain wall number one, two, and three terms in the potential.}

\subsection{Axion detection}

\begin{figure}[t!]
\begin{center}
\begin{tabular}{c}
\includegraphics[width=7cm]{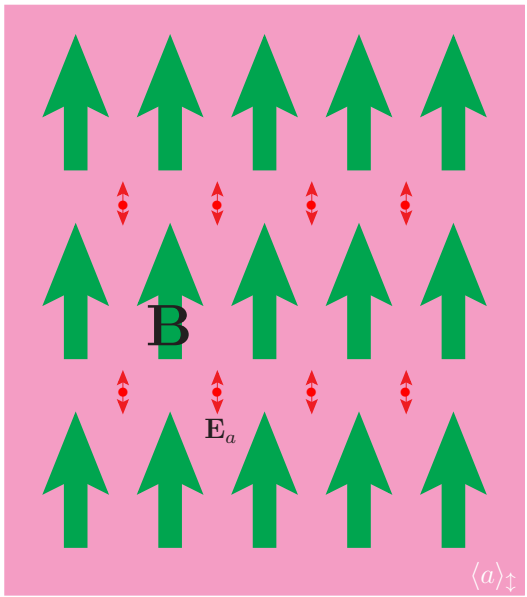}
\end{tabular}
\end{center}
\caption{The (constant) magnetic and (oscillating) electric field directions used
in the axion detection experiments. The electric field follows the oscillation
of the classical axion field.}\label{figBandE}
\end{figure}

Figure \ref{figBandE} captures the idea behind the main experimental axion dark matter
detection effort. There are two equivalent pictures describing the axion to photon conversion
in the presence of a Direct-Current (DC) magnetic field ${\bf B}$ \cite{Sikivie89,Krauss85,Jihnekim14},
briefly described here: The axion decays to two photons through the triangle anomaly.
Its lifetime, for an axion mass in the $\mu$eV range, is of order $10^{50}$~s, much
larger than the lifetime of our Universe.  This decay rate can be significantly enhanced
in the presence of a DC ${\bf B}$-field via the inverse  Primakov effect \cite{Sikivie89}.
This decay rate is additionally enhanced by the density of the final states, e.g., the
quality factor $Q$ of a resonant microwave cavity when its resonant frequency coincides
with the axion field oscillation frequency.

In the second picture the axion couples to the product of ${\bf E} \cdot {\bf B}$,
where ${\bf E}$ is an electric field.
In the presence of a DC magnetic field there is an oscillating electric field appearing
with the same frequency as the axion field.
If the DC magnetic field is of finite extent, then the oscillating ${\bf E}$-field induces
an azimuthal oscillating magnetic field due to Maxwell's equations.
If there is a resonant microwave cavity at the same boundary and with
the same resonant frequency, it then provides feedback that enhances the
oscillating ${\bf E}$-field by the quality factor of the cavity \cite{Krauss85,Jihnekim14}.
The power conversion of the axion DM to microwave photons estimated
by the two methods is the same and it is given by
\begin{equation}
P_{a \to \gamma} = g_{a \gamma \gamma} ^2
\left( \frac {\rho_a}{m_a} \right)
B_0^2 V C_j Q_L \,, \label{eq:axpower}
\end{equation}
where $Q_L$ is the cavity loaded quality factor, $C_j$ is the mode filling factor,
$g_{a \gamma \gamma}$ is the axion-photon-photon coupling constant,
$\rho_a$ is the axion DM local density, $m_a$ is the axion mass, $B_0$
is the strength of the DC magnetic field and $V$ the volume of the cavity.

The expected power conversion $P_{a \to \gamma}$ is extra-ordinarily small,
but nonetheless it can be within the present experimental capabilities for an axion
mass in the 1--20 $\mu $eV range.  Have we known the axion mass with a
1 part per million (ppm) accuracy, it would take less than a day to detect it
if axions were more than 10\% of the DM.
The main issue is that, barring the BICEP2 results \cite{Marsh14,Gondolo14},
we have no such information.  The best-suited axion DM mass is below about
1 meV all the way to about 1 $\mu$eV, spanning three orders of magnitude with
a potential line width of about 1 ppm.
Clearly, scanning the whole axion mass range will require too many steps,
and therefore the sensitivity needs to be very high at each step.

\begin{figure}
\begin{center}
\begin{tabular}{c}
\includegraphics[width=7.5cm]{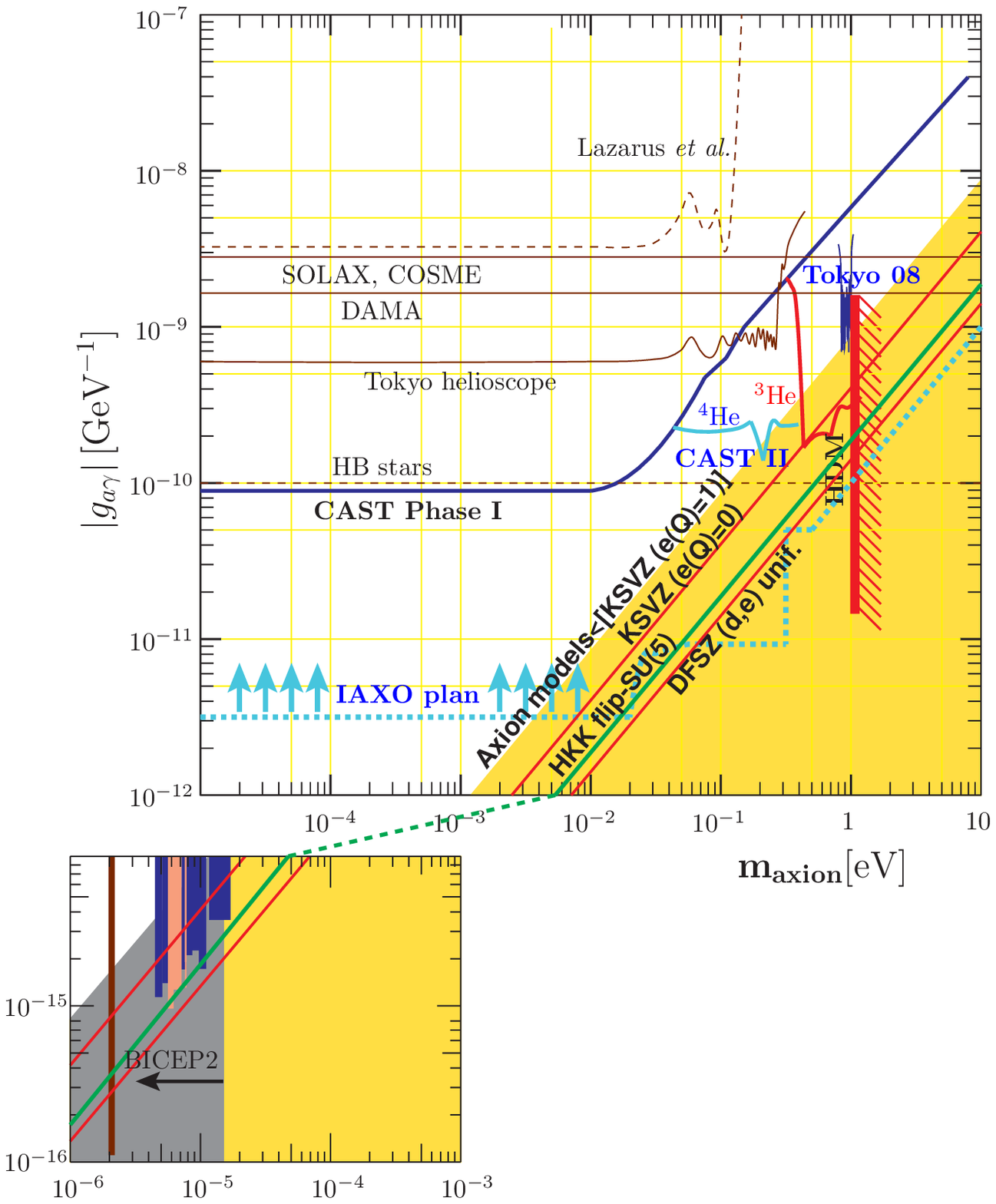}
\end{tabular}
\end{center}
 \caption{ Experimental/astrophysical limits of $g_{a\gamma}$ vs. $m_a$ \cite{RevAxionsCDM}. The experiments giving the limits are shown. The KSVZ and DFSZ lines are from Ref. \cite{Kim98}, and the string calculation is the green line \cite{Kimagg14}.  The limit of the yellow region is the largest one from Ref. \cite{Kim98}. The excluded gray region is from \cite{Marsh14,Gondolo14}, allowing a factor 5 generosity due to the domain wall annihilation problem \cite{BarrKim14}. The astrophysics lines in the bigger box represent that the regions above those lines are excluded.}\label{figExp}
\end{figure}

Furthermore, in some theoretical scenarios, the axion DM mass is not
constrained from below and can be very light, well below 1 $\mu$eV.
In addition to the microwave cavity method, which is mostly applicable
between 1-20 $\mu$eV, other methods include looking for axions emitted by
Sun's core, and astrophysical limits, as axions can provide another channel of
energy loss, significantly altering the star lifetime.  An overview of the present
experimental/astrophysical limits of the axion coupling constant vs.
the axion mass are given in Fig.~\ref{figExp}.

Looking at Eq.~(\ref{eq:axpower}), it is clear there is a number of possible
improvements one can make in this method:
(i) Increase the magnetic field value,
(ii) Increase the magnetic field volume, and
(iii) Increase the cavity quality factor.
The pioneering axion DM experiments that started in the late
1980's \cite{RBF,Florida} probed an axion DM candidate in a limited mass
region, assuming a stronger axion to photon coupling than is required by
theory by roughly two orders of magnitude.

Over a period of more than 15~years, the dominant axion dark matter experiment (ADMX),
currently located at the University of Washington and ADMX-HF located at
Yale University, have made several conceptual improvements and have improved on those limits.
The second generation ADMX experiment, owing to the development of
very low noise SQUID amplifiers just below 1~GHz \cite{Clarke10} and a number
of additional smaller developments, has reached the boundaries of a plausible axion DM
candidates.  Currently implementing a dilution refrigerator to their system is expected
to allow them to either detect or exclude an axion comprising 100\% of the DM for
masses in the range 1--20 $\mu$eV.

The new Center for Axion and Precision Physics (CAPP)  \cite{capp_site}, established
by the Institute for Basic Science  in South Korea \cite{ibs_site}, plans to
either detect or exclude an axion DM component down to the 10\% level for
a similar axion mass range.
This will be achieved by \cite{Yannis14} (a) Development of a 25\,T and then a 35\,T
solenoidal magnet compared to the currently used 8--9\,T solenoidal magnets,
(b) Substantially improving, roughly by an order of magnitude, the quality factor
of the microwave cavities in the presence of strong magnetic fields, and
(c) Constructing and running a toroidal cavity with a large volume and
a reasonable ${\bf B}$-field value so that the overall product $B^2 V$ is
an order of magnitude larger than present values.

The commonly used NbTi superconducting cable has a critical current that falls very rapidly as the magnetic field increases above 10\,T, making it unsuitable to obtain higher B-field strengths.  However, recent developments with high $T_c$ cables makes possible achieving much higher current densities at large B-field values, when they are cooled at low temperatures around 4\,K.  This is an experimental method fuelled by the energy-storage field and prototype magnets are already under development.  CAPP is collaborating with the Magnet Division of Brookhaven National Laboratory to develop a 10cm inner bore diameter capable of producing around 25\,T of magnetic field.  Preliminary tests on different high $T_c$ cables are providing encouraging results that the goal can be met.  The expected time period for this development is of order of five years, after which we develop a separate magnet with a goal of achieving 35\,T peak magnetic field, albeit with smaller inner bore diameter.
The next step would be to configure a toroidal magnetic field, optimising the use of the magnetic field as the fringe field is minimized in that geometry.  Preliminary cable testing results also point to this geometry for achieving the highest possible magnetic field values.  The time scale for this development is of order ten years.

The presently used cavities have a quality factor between 50\,K--100\,K.  It has been reported by ADMX that they are developing cavities with thin-film superconducting coatings on the vertical side walls with the goal of increasing the cavity quality factor by roughly a factor of five.  This is possible when the B-field is shaped to be aligned with the vertical wall, minimising the transverse B-field below about 100 Gauss.
Further increases of the quality factor are hindered by the top/bottom surfaces of the right-cylindrical cavity as the magnetic field angle traversing the surface is very close to 90 degrees.  Our plan to further improve upon this achievement is two-fold: First, develop a toroidal cavity where the B-field can be shaped along the cavity walls reducing the transverse B-field below the required level.  If that is possible, the quality factor can be increased by several orders of magnitude. Second, the top/bottom plates are going to be treated in a way that the B-field can penetrate it without affecting the superconducting layer on the inside of the cavity.  Again, the quality factor can potentially increase by several orders of magnitude.

The expected axion width is of order 1\,ppm, {\it i.e.} the axion quality factor is a bit better than $10^6$.  Therefore, the best one can do is to produce a cavity with the same quality factor, so the best one can expect is to gain a factor of 10 to 20 in the axion to photon power conversion.  The scanning speed goes as the square root of the quality factor since there are more steps required in order to cover all possible frequencies,  {\it i.e.}  the best one can expect to do is a scanning speed improvement factor between three and five.

BICEP2 results favour axion masses in the meV range, albeit with only 1--10\,\% of
DM composed of axions.  This fact makes it particularly difficult to detect it as
the volume of microwave cavities are particularly small and not of much practical
use at those frequencies, plus the axion DM density is very weak.
If the BICEP2 results turn out to be confirmed,\footnote{
We note that the recent Planck-dust report extrapolated to the BICEP2 field
gives the dust contribution similar to $r\approx 2$ without the dust contribution \cite{PlunckDust14}.} one could follow
a different strategy in detecting axions \cite{Asimina14}.
If the axion mass were to be found, then one could launch a dedicated
axion DM experiment within a very small axion mass range having
much higher chances of success.

\section{The cosmological constant problem and string theory}\label{sec:CC}

In order to realize the present-day cosmic acceleration with
the cosmological constant  $\Lambda$, we require that
$\Lambda$ is of the order of $H_0^2$, i.e.,
$\Lambda \approx H_0^2=(2.1332h \times 10^{-42}~{\rm GeV})^2$,
where $h \approx 0.7$.
This corresponds to the energy density
$\rho_{\Lambda} \approx \Lambda \Mp^2 \approx 10^{-120}\Mp^4$.
Even before the discovery of the present-day cosmic
acceleration, Weinberg \cite{WeinbergAnth}
put the bound on $\rho_{\Lambda}$, as
\begin{equation}
-2\times 10^{-120}\Mp^{4} \lesssim \rho_{\Lambda}\lesssim
6\times10^{-118}\Mp^{4}\,.
\label{rhoregion}
\end{equation}
The lower bound comes from the fact that the negative cosmological constant does not lead to the collapse of the Universe today. The upper bound corresponds to the requirement that
the vacuum energy does not dominate over the matter density
for redshifts $z$ larger than 1 to realize the successful structure formation.

There have been attempts to explain the very low values of
$\rho_{\Lambda}$ ranging the Weinberg bound (\ref{rhoregion}).
For example, Bousso and Polchinski \cite{Bousso} employed the 4-form
field $F_{\mu \nu \lambda \sigma}$ with the energy density
$F_{\mu \nu \lambda \sigma}F^{\mu \nu \lambda \sigma}/48=c^2/2$,
where $c$ is a constant. In the context of string theory,
there are ``electric charges'' (membranes) sourcing the 4-form field
dual to ``magnetic charges'' (5-branes). The constant $c$ can be
quantized in integer ($n$) multiples of the membrane charge $q$, such that $c=nq$.

Bousso and Polchinski introduced $J$ 4-form fields together
with $J$ membrane species with charges $q_{1},q_{2},\cdots,q_{J}$. The number $J$ can be
as large as 100 in string theory. Since the flux energy density of each charge is given by
$n_{i}^{2}q_{i}^{2}/2$, the effective cosmological constant reads
\begin{equation}
\Lambda=\Lambda_{b}+
\sum_{i=1}^{J}n_{i}^{2}q_{i}^{2}/2\,, \label{qless2}
\end{equation}
where $\Lambda_b$ is the bare cosmological constant.
For the anti de Sitter minimum with $\Lambda_b<0$,
there exist integers $n_i$ satisfying
\begin{equation}
2|\Lambda_{b}|<\sum_{i=1}^{J}n_{i}^{2}q_{i}^{2}
<2(|\Lambda_{b}|+\Delta\Lambda)\,,\label{nicon}
\end{equation}
where $\Delta\Lambda\simeq10^{-123}$ in the unit $\Mp=1$.

If we consider a $J$-dimensional grid with axes corresponding
to $n_{i}q_{i}$, the displacement of the 4-form field is given
by discrete grid points with integers $n_{i}$.
The region (\ref{nicon}) corresponds to a thin-shell characterized
by the radius $r=\sqrt{2|\Lambda_{b}|}$ and the width
$\Delta r=\Delta\Lambda/\sqrt{2|\Lambda_{b}|}$.
The volume of the thin-shell is
\begin{equation}
V_{S}=\Omega_{J-1}r^{J-1}\Delta r=
\Omega_{J-1}|2\Lambda_{b}|^{J/2-1}\Delta\Lambda\,,
\end{equation}
where $\Omega_{J-1}=2\pi^{J/2}/\Gamma(J/2)$ is the area of a unit
($J-1$)-dimensional sphere. A grid cell has a volume
$V_{C}=\prod_{i=1}^{J}q_{i}$.
There is at least one value of $\Lambda$ for $V_{C}<V_{S}$, i.e.,
\begin{equation}
\prod_{i=1}^{J}q_{i}<\frac{2\pi^{J/2}}{\Gamma(J/2)}
|2\Lambda_{b}|^{J/2-1}\Delta\Lambda\,.\label{qicon}
\end{equation}
When $J=100$, $|\Lambda_{b}|=1$ and $\Delta V=10^{-123}$ with
equal charges ($q_{i}=q$, for $i=1,2,\cdots,J$), the
condition (\ref{qicon}) is satisfied for $q<0.035$. Since the
charge $\sqrt{q}$ has the dimension of mass from Eq.~(\ref{qless2}),
this condition translates to $\sqrt{q}<0.19$ in units of \Mpt. Thus, the presence
of many 4-form fields allows the possibility of realizing a small effective cosmological constant.

The idea of Bousso and Polchinski is based on the flux energy density
originating from multiple 4-form fields.
This idea was extended to the so-called flux compactification on a Calabi-Yau
manifold in type II string theory.
In the presence of fluxes, Kachru, Kallosh, Linde and Trivedi \cite{KKLT}
first set up a supersymmetric anti de Sitter (AdS) vacuum with
all moduli fields fixed.
Then, they obtained a de Sitter vacuum by adding an anti D3-brane
in a warped geometry to lift up the AdS state.

There are hundreds of different 3-cycles on the Calabi-Yau manifold in
the flux compactification.
A macroscopic observer can view a 5-brane wrapping a 3-cycle
as a 2-brane (membrane).
The 5-brane can wrap any of these 3-cycles, which gives
rise to hundreds of different membranes in four-dimensional space-time.
The number of vacua appearing in string theory can be extremely large.
For 500 three-cycles with each cycle wrapped by up to 10 fluxes,
we have $10^{500}$ vacua.

The possible presence of such a large amount of vacua
led to the notion of so-called string landscape \cite{Susskind}.
This landscape includes so many possible configurations of local minima,
among which our Universe may correspond to one of them.
Each vacuum in the string landscape has different matter and coupling constant.
The SM is not predicted uniquely in this picture. The argument
is that we may be able to find a vacuum with
an extremely small energy density among $10^{500}$ vacua.
However, this anthropic argument depends on ``{Those packed near $\Lambda=0$ out of $10^{500}$ vacua describe particle phenomenology correctly, in particular with three chiral families and $\sin^2\theta_W=3/8$},''
otherwise the landscpe vacua differring by $\Delta\Lambda$ describe unacceptable universes.
From this reasoning, the string landscape is commented in \textbf{Table\,\ref{Tab:01}} as `not yet' established.

A general problem with the anthropic arguments is that they are often
applied to a single parameter while fixing all the others.
A parameter value that is ruled out in one case may be acceptable
if something else is changed at the same time.
In this sense, it is not clear that the anthropic arguments of $\Lambda$
provide a satisfactory answer to the cosmological constant problem.

As commented before, the DE scale may be accountable from highly suppressed
non-renormalizable terms in string-allowed discrete symmetries \cite{KimNilles14,KimJKPS14}
if the true vacuum has zero cosmological constant. In this sense, the theoretical solution toward the vanishing cosmological constant  is more difficult to solve than obtaining a tiny DE scale on top of
the vanishing cosmological constant  \cite{WeinbergCC,ZeeEgr13}.

\section{Inflation}\label{sec:inflation}

The possibility of an exponential expansion of the Universe
was known \cite{LeMatre,Sato75,Sato81,Kazanas80,Star80}
even before the influential paper of Guth \cite{inflationold} which advocates
diluting away the GUT scale monopoles \cite{Preskill79}.
For example, in the abstract of the Kazanas's paper \cite{Kazanas80}, it is stated that
``...In particular it is shown that under certain conditions this expansion law is exponential.
It is further argued that under reasonable assumptions for the mass of the associated
Higgs boson this expansion stage could last long enough to potentially account for
the observed isotropy of the universe.''
In the papers of Sato  \cite{Sato81,Sato75}, diluting away topological defects such as monopoles
and domain walls was stressed after the advent of the modern GUT model \cite{GG74,PS73}.
In the Guth's paper \cite{inflationold} it was clearly emphasized that the inflationary
paradigm can address the solutions for the homogeneous, horizon and flatness problems.

The scalar field responsible for inflation is called `inflaton'.
The inflaton field is a superposition of quanta of all possible wave lengths.
A quantum fluctuating scale inflates exponentially and after passing the horizon,
it is stretched exponentially with an almost scale-invariant
form \cite{Mukhanov81,Hawking82,Starobinsky82,GuthPi82,Bardeen83} and the frozen-scale
still inflates exponentially (see Ref.~\cite{BTW} for a review).
Different fluctuating scales go out of the horizon at different cosmic times
and their exponentially stretched scales are correlated.

After the end of inflation, the quantum fluctuations enter the horizon again
and become the sources of density perturbations.
The prediction of nearly scale-invariant primordial perturbations generated during inflation
was consistent with the  temperature anisotropies of
Cosmic Microwave Background (CMB) observed by the COBE satellite \cite{COBE92}.
The recent WMAP and the Planck data of CMB refined the temperature anisotropies to very
high accuracy \cite{WMAP9,Planck13}\footnote{From the Planck data the existence of
CDM was also confirmed (by 7\,$\sigma$ \cite{Turner14Talk})
better than any other data.}.

The observables and the constraints implied by inflation are
\begin{itemize}
\item {A sufficient inflation, requiring the large e-fold number, $N_e>70$, for
addressing horizon and flatness problems.
\item The amplitude of temperature anisotropies
$\delta T/T\simeq 10^{-5}$, for galaxy formation with CDM.
\item The spectral index of scalar perturbations $n_s\simeq 0.96$, from WMAP and Planck data.
\item The tensor-to-scalar ratio $r \lesssim 0.2$, from WMAP and Planck data.
\item The non-linear estimator of scalar non-Gaussianities for the local shape
$f_{\rm NL}^{\rm local}=2.7 \pm 5.8$ (68\% CL), from Planck data.
}
\end{itemize}

As long as the slow-roll conditions are satisfied, the single-field inflationary
scenario generally gives rise to local non-Gaussianities with
$|f_{\rm NL}^{\rm local}|$ much smaller than 1 even for most
general scalar-tensor theories with second order equations of
motion \cite{Cremi,Chen,DeTsu}.
Hence the slow-roll single-field models are consistent with the Planck
bound of non-Gaussianities.
Using the observational bounds of $n_s$ and $r$, we can distinguish
between many single-field inflationary models \cite{Planckinf,Kuro,PTP}.
For example, the self-coupling potential $V(\phi)=\lambda \phi^4/4$ \cite{Linde83}
and hybrid inflation \cite{Hybrid} with $n_s>1$ are disfavored from the data.

The amplitude of tensor perturbations is given by ${\it P}_h=2H^2/(\pi^2 \Mp^2)$,
so the detection of gravitational waves in CMB observations implies that
the energy scale of inflation is directly
known \cite{GWRus75,GWStar79,GWRuba82,GWFabri83,GWAbbott84}.
Since the B-mode polarization of CMB is generated by tensor perturbations
but not by scalar perturbations, the B-mode detection is a smoking gun
for the existence of primordial gravitational waves.

If the tensor-to-scalar ratio $r$ is smaller than the order of 0.01,
it is not easy to detect the CMB B-mode polarization.\footnote{However, the future
observations like LiteBIRD may reach this range.}
If $r$ is detected in the range  $r>0.05$,
then the energy scale during inflation corresponds to the GUT scale.
The great interest in the announcement of $r\sim 0.16$ from the BICEP2 group
\cite{Larger}
is because of the implication that the Universe once passed the vacuum energy scale
of $10^{16}\,\gev$. Even though the GUT scale \Mg~ is humongous from our TeV scale Standard Model,
it is  tiny from the point of gravity scale, the Planck mass \Mpt.
Because of the micro density perturbation, the vacuum energy at the scale
$(10^{16}\,\gev)^4$ leads to $r\sim {\rm O}(0.1)$. This phenomenon of the GUT scale
energy density during inflation is usually parametrized by chaotic inflation with
the potential $V(\phi)=\frac12 m^2\phi^2$ \cite{Linde83}.

\begin{figure}
\begin{center}
\includegraphics[width=7cm]{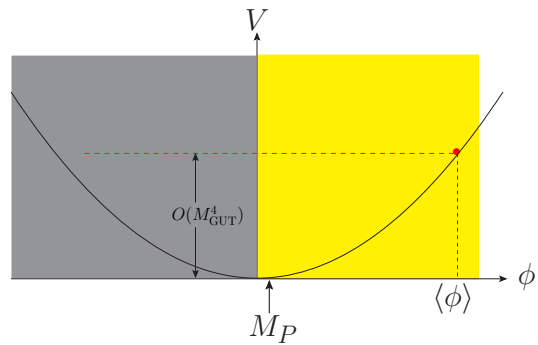}
\end{center}
\caption{Chaotic inflation with the potential $V(\phi)=m^2\phi^2/2$ \cite{Linde83}. The red bullet is the Lyth bound which is far above $M_{\rm P}$.}
\label{figChaotic2}
\end{figure}

\begin{figure}
\begin{center}
\includegraphics[width=7.5cm]{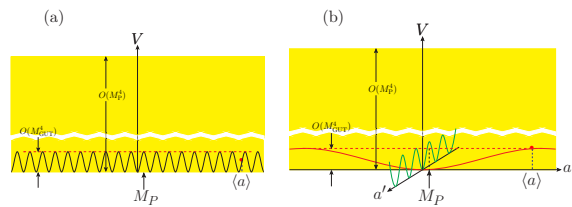}
\end{center}
\caption{ (a) The natural inflation of \cite{Freese90}; (b) The KNP inflation \cite{KNP05}. The red bullet in the 2-flation is an O(1) value of $\langle a\rangle/f_a$.}
\label{figNatKNP}
\end{figure}

If a large $r$ of order 0.2 is detected, the field value in the $\phi^2$ chaotic inflation
is bounded from below, {\it i.e.} $\langle\phi\rangle>15\Mp$,
which is known as the `Lyth bound' \cite{Lyth97}.
This situation is shown in Fig.~\ref{figChaotic2}, where the energy density
at the inflationary epoch is the GUT scale.
The field value  $\langle\phi\rangle> 15\Mp$
is trans-Planckian and the energy density at \Mpt~ is tiny.
So, one needs a fine-tuning in the $\phi^2$ chaotic inflation.
Introducing a confining force at a GUT scale, a heavy axion for the
inflaton with a potential bounded from above
was proposed \cite{Freese90}, which is called {\it natural inflation}.
In this scenario, the energy density has the upper bound of order \Mg $^4$
as shown in Fig.~\ref{figNatKNP}\,(a).
One period of the inflaton in this case is of order \Mg, and hence
the Lyth bound is violated.
To remedy this, two confining forces are introduced with two heavy
axions with the resulting potential \cite{KNP05},
\dis{
V= &-\Lambda_1^4\cos\left(\alpha\frac{a_1}{F_1}
+\beta\frac{a_2}{F_2} \right)\\& - \Lambda_2^4
\cos\left(\gamma\frac{a_1}{F_1} +\delta\frac{a_2}{F_2} \right)  +{\rm constant}\,,
}
where $\alpha,\beta,\gamma$, and $\delta$ are determined by the corresponding
PQ symmetries of two heavy axions $a_1$ and $a_2$.
Even though we allow O(1) couplings, the GUT mass scales can lead to \Mpt~
with the probability of $\sim$1\,\%. With mass parameters of 50\Mg,
we would obtain 50\Mpt with the probability of $\sim$1\,\%.
This is the Kim-Nilles-Peloso 2-flation model  \cite{KNP05}.
The probability of the 2-flation model with a large decay constant,
\ie $f_\phi >15\Mp$  to occur as shown in Fig. \ref{figNatKNP}\,(b), is about 1\%.
The green-potential in Fig.~\ref{figNatKNP}\,(b) is the other heavy axion potential.
It can be generalized to N-flation \cite{Nflation}.

The axionic topological defects in the anthropic window \cite{Pi84,Tegmark06} can be
diluted away if inflation occurs below the anthropic window scale.
With the GUT scale energy density during inflation, however, this dilution mechanism
does not work. With the GUT energy scale inflation as implied by the BICEP2 \cite{Larger},
it could have pinned down to $f_a\sim 10^{11\,}\gev$ \cite{Marsh14,Gondolo14},
using the numerical calculation of radiating axions from axionic string-wall system \cite{Kawasaki12}.
In the numerical calculation, the Vilenkin-Everett mechanism \cite{Vilenkin82} of
erasing the horizon scale string has not been taken into account.
In addition, the hidden-sector confining force can erase horizon scale
axionic strings such that the QCD axion domain wall is not a serious
cosmological problem \cite{BarrKim14}.
The hidden-sector solution needs the hidden-sector domain-wall number
of $N_h=1$, which is possible in string compactification with an anomalous U(1) \cite{KimDW14}.

In addition to pinning down the upper bound on $f_a$, the GUT scale inflation provokes
a question, ``What is the symmetry which naturally satisfies the Lyth bound \cite{Lyth97}?"
 Lyth considered this problem with respect to the $\eta$ parameter \cite{Lyth14bicep}.
But, there exists a more fundamental question. In an ultra-violet completed theory
such as string theory, every parameter is calculable.
If we consider the $\phi^2$ chaotic inflation of Fig.~\ref{figChaotic2}, there is a question,
``Why do we neglect other terms?" In string theory, only discrete symmetries are permitted
by the compactification process. For example, a term $\phi^{104}/\Mp^{100}$ can
be possible if the discrete symmetries allow it. But with the trans-Planckian value,
for example $\langle \phi \rangle \sim 31$, the coefficient must be tuned to 1 out of $10^{127}$,
which is as bad as the cosmological constant problem.

Fortunately, there is another way for inflation to occur.
We must choose the hilltop inflation, but sacrificing the single-field inflaton.
It is not so bad in view of the fact that the 2-flation model already introduced
two axions in the inflaton sector.
Then, the inflating region is near origin such that the minimum at $f_{\rm DE}$ is
far away from the origin. In the region $[0,f_{\rm DE}]$ the vacuum energy is of order \Mg$^4$.
This can be obtained from the condition on the quantum numbers of the assumed discrete
symmetry \cite{KimHilltop14}. The inflaton rolls in the yellow region in Fig.~\ref{figHilltop}
where the inflaton takes a green curve in the two-inflatons space.

\begin{figure}
\begin{center}
\includegraphics[width=7.5cm]{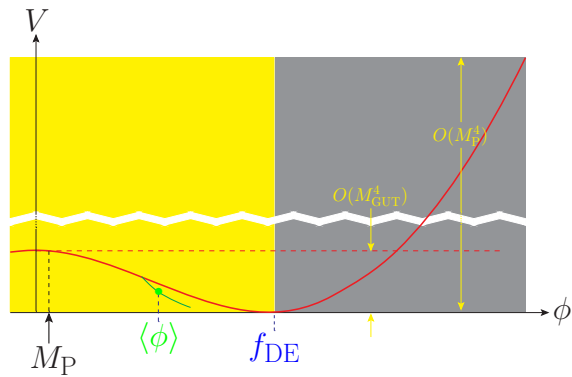}
\end{center}
\caption{The Hilltop inflation \cite{Freese90}. The green bullet is the point whose effects are observed by the BICEP2 group.} \label{fig:NatKNP}
\end{figure}

\section{Discussion}\label{sec:discussion}

After the discovery of the fundamental Brout-Englert-Higgs boson, which is represented as
$H_u$ and $H_d\,(=H_u^\dagger\,\rm ~in~ non-SUSY~case)$,
we reviewed the cosmological role of spin-0 bosons.
This finding hints the possibility of numerous spin-0 bosons ($\phi$)
at the GUT scale. Spin-0 bosons at the GUT scale of the canonical
dimension 1 can have more important effects to low energy physics
compared to those of spin-$\frac12$ fermions of the canonical
dimension $\frac32$ (Dirac fermions $\psi,\overline{\psi}$ for example)
at the GUT scale. For example, the spin-0 contribution
\dis{
\frac{\phi^{2n}}{\Mp^{2n+2m-4}}(H_d H_u)^m \label{eq:bosonHH}
}
dominates the fermion contribution $\frac{\psi^n\overline{\psi}\,^n}{\Mp^{3n +2m-4 }}(H_d H_u)^m$ for $n,m\ge 1$.
In addition, the existence of fundamental spin-0 bosons at the GUT scale may be extended to a larger symmetry: supersymmetric GUTs, or minimal supersymmetric Standard Models from string compactification. The interactions of the singlet fields only can take a SUSY superpotential, for example with GUT scale singlets $\phi$ and trans-Planckian singlets $\Phi$ for simplicity \cite{KimHilltop14}
\dis{
W=\sum_i \frac{\phi^{a_i}}{\Mp^{a_i+\ell_i-3}} \Phi^{\ell_i}.\label{eq:Winf}
 }

The rationale leading to the forms of Eqs. (\ref{eq:bosonHH}) and  (\ref{eq:Winf}) are the discrete symmetries obtained from string compactification \cite{KimNilles14},\footnote{See, also, the discrete gauge symmetries in the field theory language \cite{KraussW89}.} which guarantees the absence of gravity obstruction of discrete symmetries, for example via wormholes \cite{Kim13worm}. The form of the interactionis  (\ref{eq:Winf}) can lead to inflation with trans-Planckian decay constant with a multi-field hilltop potential, {\it i.e.} {\bf BCM2}.  The form of the interactionis  (\ref{eq:bosonHH}) can lead to QCD axion via {\bf BCM1}, and the DE scale via   {\bf CCtmp}.  The fundamental scalars at the TeV, GUT and trans-Planckian scales allow all scenarios presented in Subsec. \ref{subsec:spin0cosm}. These are worked out on top of vanishing cosmological constant, which is assumed in any particle physics models. At present, we do not have any persuasive hint toward a theoretical solution of the vanishing cosmological constant. Any theory for the vanishing cosmological constant must satisfy the requirements of particle phenomenology we used in this review.

The fundamental scalars may be detectable if their couplings to gluons are  appreciable.  The front runner in the search of fundamental scalars hinting high energy (GUT or intermediate) scales is the QCD axion which couples to the gluon anomaly.

\section*{Acknowledgement}

  JEK is supported in part by the National Research Foundation (NRF) grant funded by the Korean Government (MEST) (No. 2005-0093841) and  by the IBS (IBS-R017-D1-2014-a00), YKS is supported by the IBS (IBS-R017-D1-2014-a00), and ST is supported by the Grant-in-Aid for Scientific Research from JSPS (No. 24540286).

\section*{Disclosure/Conflict-of-Interest Statement}
The authors declare that the research was conducted in the absence of any commercial or financial relationships that could be construed as a potential conflict of interest.


\end{document}